\documentclass[cameraready]{Interspeech}

\title{LightBeam: An Accurate and Memory-Efficient CTC Decoder for Speech Neuroprostheses}

\author[affiliation={1,2}, orcid=0000-0001-9491-0547, equalcontribution, correspondingauthor]{Ebrahim}{Feghhi}
\author[affiliation={2}, orcid=0009-0006-3896-5918, equalcontribution]{Junlin}{Hu}
\author[affiliation={1,2}, orcid=0009-0000-9521-2185]{Nima}{Hadidi}
\author[affiliation={1,2,3}, orcid=0000-0002-9298-0143]{Jonathan C.}{Kao}

\address{
    $^1$ Neuroscience Interdepartmental Program\\
    $^2$ Department of Electrical \& Computer Engineering\\
    $^3$ Department of Computer Science \\
    University of California Los Angeles
}

\email{efeghhi@gmail.com}


\usepackage{comment}

\begin{document}

\maketitle

\begin{abstract}
A promising pathway for restoring communication in patients with dysarthria and anarthria is speech neuroprostheses, which directly decode speech from cortical neural activity. Two benchmarks, Brain-to-Text '24 and '25, released intracranial recordings from patients with dysarthria along with a baseline algorithm trained with Connectionist Temporal Classification (CTC). Despite significant innovation on these benchmarks, all leading published prior work relies on a WFST-based CTC decoder that requires ${\sim}$320 GB of RAM. These memory requirements limit accessibility for both patients and researchers. Here, we propose \textit{LightBeam}, a non-WFST based CTC decoder that requires only ${\sim}$10 GB of RAM and achieves state-of-the-art performance on both benchmarks. \textit{LightBeam} achieves this by integrating an LLM into the beam-search process via delayed fusion, obviating the prior need for using a large N-gram LM. \textit{LightBeam} is implemented in Python and is open-source.
\end{abstract}

\section{Introduction}
Conditions such as amyotrophic lateral sclerosis (ALS), cerebral palsy, or brainstem stroke can lead to dysarthria and anarthria. Dysarthria and anarthria are speech production disorders in which the neural activity generating speech plans in the motor cortex remains intact, yet the transmission of these motor plans to the vocal tract or the vocal tract musculature itself is impaired \cite{Silva2024-hp}. These impairments manifest as slow or slurred speech in dysarthria, and a complete loss of the ability to speak in anarthria. Speech neuroprostheses offer a promising path toward restoring communication for patients with these disorders, bypassing the vocal tract by decoding speech directly from neural activity \cite{Willett2023-an,Chen2024-hk,Card2024-an,Littlejohn2025-hw}.

Two benchmarks, Brain-to-Text '24 (B2T '24; \cite{Willett2024-ki}) and '25 (B2T '25; \cite{card2025braintotext}), have accelerated the development of speech neuroprostheses. Each competition provided intracranial electrode recordings from the speech motor cortex of a distinct participant with ALS while they attempted to repeat text prompts. Along with the neural dataset, the organizers released a baseline algorithm consisting of two components. The first component, the CTC encoder, converts neural activity into phoneme-level logits using a Gated Recurrent Unit (GRU) architecture trained with CTC loss \cite{GravesUnknown-ig}. The second component, the CTC decoder, converts logits into words by running beam search on a weighted finite-state transducer (WFST) graph integrated with a large 5-gram LM. Since LLMs cannot be represented as WFST graphs, an LLM (OPT 6.7B; \cite{Zhang2022-fg}) rescores beams at the end of the trial. This CTC decoder requires ${\sim}$320 GB of RAM. 

While significant progress has been made on these benchmarks, existing top-performing published algorithms rely on the memory-intensive baseline WFST-based CTC decoder \cite{Jia2025-uz, feghhi2025time, zhang2026decoding}. These memory requirements present a bottleneck for deployment on local, resource-constrained devices. Local deployment is ideal as it preserves user privacy, minimizes transmission latency, and maximizes accessibility for patients and the broader research community. 


In this study, we present \textit{LightBeam}, an accurate and memory-efficient CTC decoder for speech neuroprostheses. The core insight behind \textit{LightBeam} is that rather than only using an LLM for second-pass rescoring, we integrate it within the beam search process (i.e. first-pass decoding)  at fixed time intervals via delayed fusion \cite{Hori2025-hh}. By doing so, we reduce the computational burden of preserving likely beam hypotheses on the N-gram LM, and replace the large 5-gram LM with a relatively small 4-gram LM. \textit{LightBeam} reduces memory requirements from ${\sim}$320 GB to ${\sim}$10 GB of RAM. Furthermore, it performs significantly better than the baseline WFST-based decoder on both benchmarks when paired with either the baseline GRU or the time-masked Transformer encoder introduced by Feghhi et al. ~\cite{feghhi2025time}. When applying generative error correction, \textit{LightBeam} achieves state-of-the-art (SOTA) performance compared to published results by Zhang et al. ~\cite{zhang2026decoding} (that also use generative error correction) on both benchmarks. While \textit{LightBeam} requires making more frequent calls to an LLM, real-time factor (RTF) values are well below 1, so it remains viable for clinical use. Furthermore, \textit{LightBeam} is open-source and implemented in Python, making it easily modifiable and deployable in modern research environments, with code provided \href{https://github.com/ebrahimfeghhi/brainaudio}{here}.

\section{Related Works}
Prior work attempted to entirely bypass the WFST-based decoder by projecting neural activity into the embedding space of an LLM, and fine-tuning the LLM to directly generate text \cite{Feng2024-do, zhang2026decoding}. While this multi-modal LLM approach has been successful in automatic speech recognition (ASR) \cite{Jia2025-uz}, it performs substantially worse than the WFST decoder in this context, likely due to small neural dataset sizes \cite{zhang2026decoding}. One reason researchers prioritized building multi-modal LLMs over integrating them into the beam-search process may be that the majority of open-source CTC toolkits only support N-gram LM integration  \cite{kensho2021pyctcdecode, galvez2023gpu, Grigoryan2025-fh}. While the Flashlight decoder on PyTorch does support neural network integration, it is CPU-based and thus cannot support LLM inference \cite{Kahn2022-qs}. Hori et al. \cite{Hori2025-hh} were the first to demonstrate that LLMs could be effectively used in first-pass decoding without additional projection layers; however, they released no open-source implementation. Furthermore, their results are shown in the context of ASR with encoders that output sub-word tokens and are already accurate without external LM integration. By contrast, CTC encoders for neural speech decoding typically output phonemes and heavily rely on external LM guidance due to small dataset sizes. 

\section{Methods}

\subsection{Dataset}   
\label{sec:dataset}
Our analyses utilize the neural datasets from the Brain-to-Text '24 (B2T '24) and '25 (B2T '25) benchmarks \cite{Willett2023-an, Card2024-an}. The B2T '24 dataset consists of 12,100 sentences recorded from participant T12 over 25 sessions spanning 4 months, using 128-electrode Utah arrays implanted in the ventral motor cortex. The B2T '25 dataset includes 10,948 sentences from participant T15 across 45 sessions over 20 months, recorded with 256 electrodes in the ventral motor cortex. Across both datasets, neural features consist of thresholded spike counts and spiking-band power \cite{Nason2020-lp}. These signals were binned in 20 ms windows and z-scored within blocks. Additional details regarding these datasets, including train, val, and test split construction are included in Supplementary Section A as well as the original studies. 

\subsection{Encoder models}
\label{sec:ctc_encoder}
We trained the baseline GRU encoders using the hyperparameters and codebases provided by the B2T '24 and '25 benchmark organizers. The B2T '24 GRU was bidirectional (non-streaming), while the B2T '25 GRU was unidirectional. Additionally, we evaluated \textit{LightBeam} using the time-masked Transformer, which outperforms the baseline GRU on B2T '24 \cite{feghhi2025time} and achieves state-of-the-art published results when combined with self-supervised pretraining \cite{zhang2026decoding}. For B2T '24, we log-transformed the neural data following Feghhi et al. \cite{feghhi2025time}; for B2T '25, we utilized the z-scored data provided by the organizers without further transformation. To address the challenge of the unbounded KV-cache in self-attention—which complicates clinical deployment—we trained a causal Transformer with limited left context (see Supplementary Section C). For all architectures, a final linear layer produced 41 token classes: 39 phonemes, a CTC blank token, and a space token indicating word boundaries. All encoders were trained using CTC loss.

\subsection{WFST decoder}
\label{sec:wfst}
We evaluated \textit{LightBeam} against the baseline WFST-based decoder described by Willett et al. \cite{Willett2023-an}. This approach represents the logit-to-token mapping ($\mathcal{T}$), the phoneme-to-word mapping (lexicon, $\mathcal{L}$), and the N-gram LM (grammar, $\mathcal{G}$) as individual WFSTs, which are pre-compiled into a single, unified graph ($\mathcal{T} \circ \mathcal{L} \circ \mathcal{G}$). During inference, beam search is performed on this graph with a pruned 5-gram LM to generate a word lattice, which is then subsequently rescored using an un-pruned version of the same 5-gram LM. In a final rescoring stage, the top 100  candidate beams are extracted and scored using OPT 6.7B \cite{Zhang2022-fg}. These LLM scores are combined with the un-pruned 5-gram LM scores and scaled CTC encoder log-probabilities to determine the final output. We used the exact decoding hyperparameters specified by Willett et al. \cite{Willett2023-an} and Card et al. \cite{Card2024-an} for the B2T '24 and '25 benchmarks, respectively (Supplementary Section D).

\begin{algorithm}[t]
\scriptsize 
\SetAlgoVlined 
\SetKwInOut{Input}{Input }
\SetKwInOut{Output}{Output}
\SetKwComment{tcp}{}{}
\SetInd{0.33em}{0.66em}

\providecommand\mycommfont[1]{\footnotesize\textcolor{green!60!black}{#1}}
\renewcommand\mycommfont[1]{\footnotesize\textcolor{green!60!black}{#1}}
\definecolor{royal_blue}{RGB}{20, 80, 200}
\SetCommentSty{mycommfont}

\caption{LightBeam}
\label{alg:LightBeam}


 \tcp{\scriptsize Scaled log-probs $\in \mathbb{R}^{T \times k}$}                                                                                                       
  $D$ = $\alpha \cdot$ \texttt{Log\_Softmax(encoder\_logits)}\;
  \BlankLine                                                                                                                                                             
  \texttt{Beams} = \texttt{InitBeams}($k$, $o$, $|\mathcal{V}|$)\;
  \texttt{VecLex} = \texttt{InitVectorizedLexicon}($k$, $|\mathcal{V}|$, $\mathcal{L}$, $\mathcal{T}$)\;                                                                 
  \BlankLine      
  \tcp{\scriptsize Maintain state within word prefix for each beam, $\in \mathbb{R}^{k}$}
  \texttt{prefix\_state} = \texttt{VecLex.Init\_State()}\;

  \BlankLine
  \For{$t \leftarrow 0$ \KwTo $T - 1$}{

      \BlankLine
      \tcp{\scriptsize Extend $k$ beams with $D[t]$ to form current scores $\in \mathbb{R}^{k \times |\mathcal{V}|}$}
      \texttt{scores} = \texttt{Fan\_Out\_Beams(Beams.scores, D[t])}\;

      \texttt{scores} = \texttt{Add\_Token\_Word\_Insertion\_Bonus(scores)}\;

      \BlankLine
      \tcp{\scriptsize Lexicon mask $\in \mathbb{R}^{k \times |\mathcal{V}|}$ is true for valid extensions. Repeated and blank tokens are always valid.}
      \texttt{lex\_mask} = \texttt{VecLex.Get\_Mask(prefix\_state, \\ Beams})\;
      \texttt{scores[$\neg$lex\_mask]} = $-\infty$\;

      \BlankLine
      \tcp{\scriptsize Apply top-k selection and threshold pruning.}
      \texttt{scores, indices} = \texttt{select\_top\_k(scores)}\;
      \texttt{scores[scores < max(scores) - $\theta$]} = $-\infty$\;

      \BlankLine
      \tcp{\scriptsize Update beam states after pruning.}
      \texttt{Beams.update(scores, indices)}\;

      \BlankLine
      \tcp{\scriptsize Update prefix states for beams extending with non-blank, non-repeating tokens.}
      \texttt{prefix\_state} = \texttt{VecLex.Update(prefix\_state, \\ Beams)}\;
      \BlankLine

      \tcp{\scriptsize Apply shallow fusion w/ N-gram LM, beam scores update in place.}
      \texttt{ApplyNGram(Beams, \texttt{N-gram LM}, $\lambda$)}; \tcp{\scriptsize See Section \ref{sec:n_gram_lm_lightbeam}}
      \BlankLine
      \tcp{\scriptsize Apply delayed fusion with LLM at specified intervals}
      \If{$(t \bmod \textnormal{llm\_rescore\_interval} == 0) \wedge (t > 0)$}{
          \texttt{ApplyLLM(Beams, \texttt{LLM})}; \tcp{\scriptsize See Section \ref{sec:llm_lightbeam}}
      }
  }
  \texttt{ApplyFinalLLM(Beams, \texttt{LLM})}\;
\end{algorithm}

\begin{table*}[t]
  \centering
  \caption{Combined performance and computational efficiency on B2T '24 and '25 benchmark test sets when using the baseline GRU encoder. Peak RAM and VRAM are reported in GB. Processing speed is measured in RTF using an GeForce RTX 5090 GPU. Lower is better for all metrics. Mean $\pm$ SEM across $n=10$ seeds. WFST-based (Orig.) is a single seed result provided by original authors, and re-implementation (re-impl) values are generated in this study. ($^*$ sig. better than Orig., $^\dagger$ sig. better than Re-impl.)}
  \label{tab:combined_results}
  \setlength{\tabcolsep}{6.5pt} 
  \begin{tabular}{l | cccc | ccccc}
    \toprule
    \multirow{2}{*}{\textbf{Method}} & \multicolumn{4}{c|}{\textbf{B2T '24}} & \multicolumn{5}{c}{\textbf{B2T '25}} \\
    \cmidrule(lr){2-5} \cmidrule(lr){6-10}
    & \textbf{WER} & \textbf{RAM} & \textbf{VRAM} & \textbf{RTF} & \textbf{Pub. WER} & \textbf{Priv. WER} & \textbf{RAM} & \textbf{VRAM} & \textbf{RTF} \\
    \midrule
    WFST-Based (Orig.) & 9.76 & - & - & - & 6.67 & 7.00 & - & - & - \\
    WFST-Based (Re-impl.) & 9.71 \scriptsize{$\pm$} 0.08 & 322.9 & 9.0 & \textbf{0.05} & 6.31 \scriptsize{$\pm$ 0.12} & 6.72 \scriptsize{$\pm$} 0.05 & 317.8 & 18.1 & \textbf{0.01} \\
    \textbf{LightBeam (Ours)} & \textbf{9.37$^{*\dagger}$ \scriptsize{$\pm$ 0.08}} & \textbf{9.4} & \textbf{5.6} & 0.14 & \textbf{5.77$^{*\dagger}$\scriptsize{$\pm$ 0.12}} & \textbf{6.47$^{*\dagger}$\scriptsize{$\pm$ 0.05}} & \textbf{11.6} & \textbf{5.8} & 0.06 \\
    \bottomrule
  \end{tabular}
\end{table*}

\subsection{LightBeam}
\label{sec:LightBeam}
\textit{LightBeam} is adapted from the GPU-accelerated \textit{FlexCTC} decoder \cite{Grigoryan2025-fh}. Its operational flow is detailed in Algorithm ~\ref{alg:LightBeam}, with all variables for the algorithm defined in this section or Section ~\ref{sec:wfst}. Hyperparameter values as well as additional technical details are specified in Supplementary Section E. First, the logits from the CTC encoder are transformed into a log-probability and multiplied by an acoustic scale. At each encoder frame $t$, the $k$ beams are expanded by adding the scaled CTC encoder log-probabilities to generate $k \times |\mathcal{V}|$ candidates. We apply a token extension bonus $\beta$ to non-repeating phoneme tokens and a word insertion bonus $\gamma$ to space tokens. To enable GPU parallelization across beams, we represent the lexicon as a state transition table $\mathcal{T}$ rather than a trie. Each row in $\mathcal{T}$ represents a valid word prefix state, and entry $\mathcal{T}_{i,j}$ stores the resulting prefix state when observing token $j$ at prefix state $i$, with a designated ``sink state'' for out-of-vocabulary transitions. Candidates are pruned if they violate lexical constraints (i.e. enter the sink state) or fall below a threshold $\theta$ from the score of the top beam.

\subsubsection{Shallow fusion and homophone trackng with N-gram LM}
\label{sec:n_gram_lm_lightbeam}
Phoneme sequences can map onto multiple words, known as homophones. Selecting the most likely homophone fails to account for future context; for instance, because ``\textit{There}'' is more likely than its homophone ``\textit{They're}'', the sentence ``\textit{They're happy.}'' is not possible to produce under a greedy approach. We therefore implement a nested beam search such that each beam maintains its own top $o$ orthographic (i.e., word-level) beams. Upon detecting a word boundary, the $o$ orthographic beams are expanded with the candidate homophones, scored with the N-gram LM on CPU, followed by top-$o$ selection. Orthographic beams whose scores are below a threshold, $\lambda$, from the top-scoring beam are further pruned. The current score for beam $i$, $S_{i}$, is updated as:

\begin{equation}
\label{eqn:n_gram}
S_{i} \leftarrow S_{i} + \omega \cdot [\log P_{\text{ngram}}(o_{new}) - \log P_{\text{ngram}}(o_{prev})]
\end{equation}

where $\omega$ is the LM weight, $o_{prev}$ is the previous most likely orthographic beam, and $o_{new}$ is the most likely orthographic beam after integrating the incoming homophones. We use the 4.9 GB version of the 4-gram LM trained on 8.6 billion words with a 100,000 word vocabulary provided by ImagineVille project \cite{Vertanen2019}. 

\subsubsection{Delayed Fusion with LLM}
\label{sec:llm_lightbeam}
At fixed time intervals ($1$ s and for B2T '24, $1.25$ s for B2T '25), an LLM (Llama 3.2 1B Base; \cite{Grattafiori2024-zd}, denoted as $\mathcal{N}$) rescores all (up to $k \times o$) unique orthographic beams in batches using a GPU. We then replace the N-gram scores with the LLM scores:
\begin{equation}
\label{eqn:LLM}
S_{i} \leftarrow S_{i} + \phi \cdot \log P_{\text{LLM}}(o'_{new}) - \omega \cdot \log P_{\text{ngram}}(o_{new})
\end{equation}
where $\phi$ is the LLM weight and $o'_{new}$ is the LLM-preferred hypothesis. At the end of the trial, the score of each beam is updated with its most likely end-of-sentence punctuation according to the LLM. Prior to delayed fusion, Llama is fine-tuned on a next-word prediction task using the combined training and validation sentences across both datasets to adapt it to the participant's text distribution (Supplementary Section F). 

\subsection{Generative error correction}
\label{sec:generative}
Following prior work \cite{Benster2024-eh, Jia2025-uz, feghhi2025time, zhang2026decoding}, we apply generative error correction by producing 10 candidate sentences using 10 random seeds of the same encoder model, and fine-tuning an LLM to produce the ground-truth sentence given these 10 candidates. We follow the same procedure as Feghhi et al. \cite{feghhi2025time}, with Llama 3.1 8B Instruct used for generative error correction. Refer to Supplementary Section G for more details. 

\section{Results}

\begin{table}[b!]
  \caption{Word Error Rate (\%) comparison on the B2T '24 (WER) and B2T '25 (Pub. and Priv. WER) benchmarks when using the causal time-masked Transformer with bounded KV cache. Mean $\pm$ SEM over $n=10$ seeds. ($^\dagger$ sig. better than WFST decoder).}
  \label{tab:transformer}
  \centering
  \footnotesize
  \setlength{\tabcolsep}{4.5pt} 
  \begin{tabular}{l | cc | ccc}
    \toprule
    \multirow{2}{*}{\textbf{Method}} & \multicolumn{2}{c|}{\textbf{B2T '24}} & \multicolumn{3}{c}{\textbf{B2T '25}} \\
    \cmidrule(lr){2-3} \cmidrule(lr){4-6}
    & \textbf{WER} & \textbf{RTF} & \textbf{Pub. WER} & \textbf{Priv. WER} & \textbf{RTF} \\
    \midrule
    WFST-Based & 8.36{\scriptsize$\pm$.12} & 0.02& 4.95{\scriptsize$\pm$.26} & 4.81{\scriptsize$\pm$.13} & 0.01\\
    \textbf{LightBeam} & \textbf{8.08$^\dagger${\scriptsize$\pm$.12}} & 0.10 & \textbf{4.44$^\dagger${\scriptsize$\pm$.21}} & \textbf{4.59$^\dagger${\scriptsize$\pm$.14}} &  0.05 \\
    \bottomrule
  \end{tabular}
\end{table}

\subsection{Evaluation using baseline GRU encoder}
We first utilized the baseline GRU encoder (Section \ref{sec:ctc_encoder}) to evaluate the WFST decoder (Section \ref{sec:wfst}) and \textit{LightBeam} (Section \ref{sec:LightBeam}). The GRU encoder serves as a rigorous test-bed, as the WFST decoder hyperparameters were previously tuned for its specific logit distribution. As shown in Table~\ref{tab:combined_results}, \textit{LightBeam} consistently improves Word Error Rate (WER) across both benchmarks (notably, B2T '25 reports both public and private WER).

To rigorously assess these gains, we conducted a two-sided one-sample t-test comparing \textit{LightBeam}'s performance across 10 encoder seeds against the official baseline (WFST-Based Orig.). All statistical tests in this study are two-sided. \textit{LightBeam} achieved statistically significant improvements ($p < 0.05$) on both datasets. Furthermore, we reproduced results using the WFST decoder across the same 10 seeds; a paired t-test confirmed that \textit{LightBeam} significantly outperformed this re-implementation (WFST-Based Re-impl., $p < 0.05$).

Regarding resource efficiency (Table \ref{tab:combined_results}), \textit{LightBeam} achieved a dramatic reduction in peak RAM consumption from $\sim$320 GB to $\sim$10 GB, while reducing peak VRAM requirements by 1.6$\times$ and 3.1$\times$ for the B2T '24 and '25 benchmarks, respectively. Finally, although \textit{LightBeam}’s mean RTF is higher than the WFST baseline due to more frequent application of the LLM, the average RTF remains well below 1. Critically, the maximum RTF observed across all test trials remained strictly under 1 on a GeForce RTX 5090 GPU, ensuring suitability for real-time decoding (Supplementary Section H).

\subsection{Generalization to time-masked Transformer encoder}
\begin{figure}[t!]
  \centering
  \includegraphics[width=\linewidth]{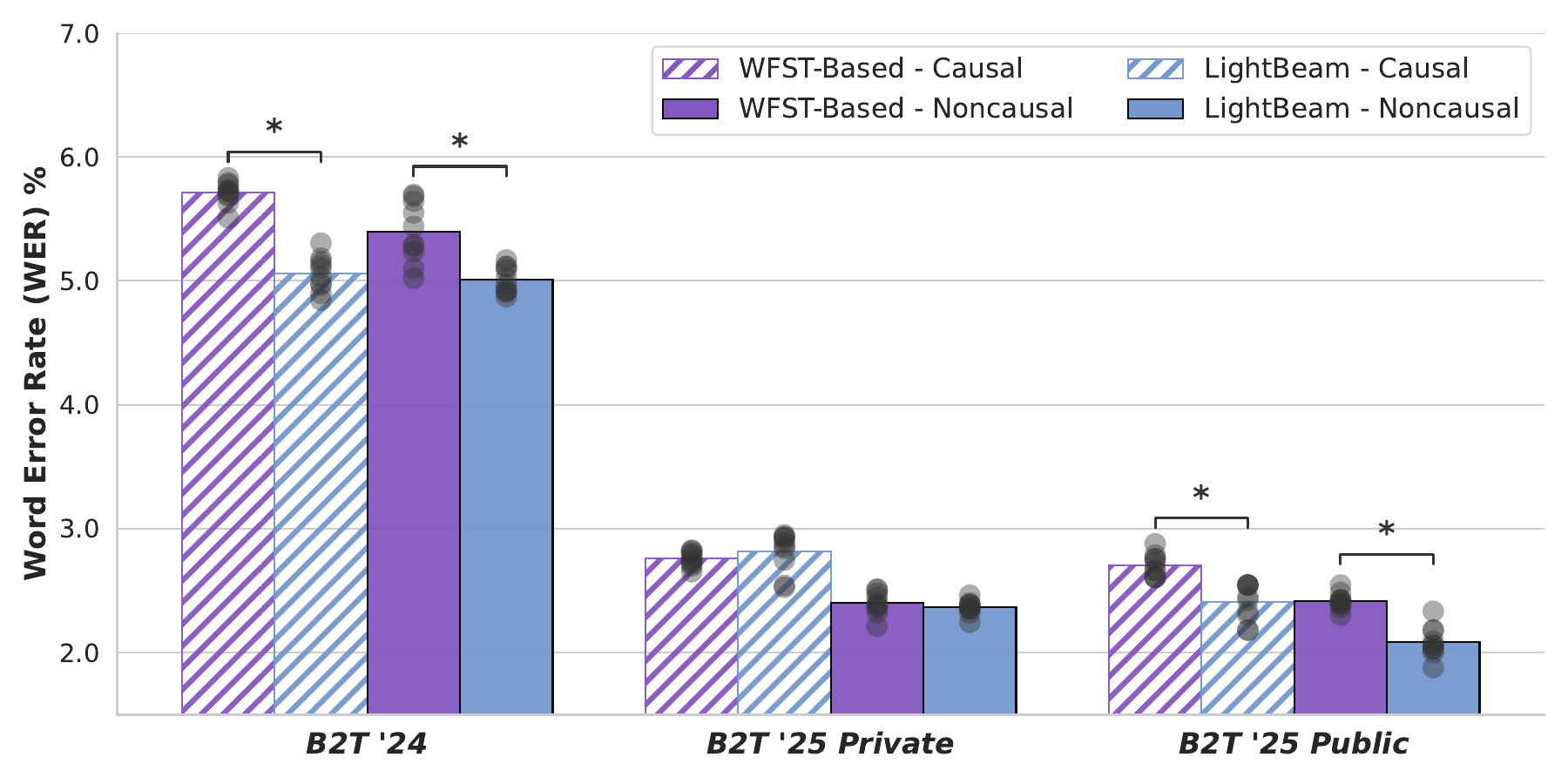}
  \caption{\textit{LightBeam} continues to match or outperform baseline WFST decoder across all benchmarks when both methods are paired with generative error correction. Results are shown with time-masked Transformer. Asterisks indicate significant improvement when performing paired t-test across $n=10$ seeds.}
  \label{fig:generative_error_corection}
\end{figure}

To evaluate whether \textit{LightBeam} is effective with other encoders, we evaluated its performance using a causal time-masked Transformer encoder with a bounded KV cache (Section \ref{sec:ctc_encoder}). No additional hyperparameter tuning was done for the WFST decoder or \textit{LightBeam} using the Transformer encoder. As shown in Table~\ref{tab:transformer}, \textit{LightBeam} also consistently outperforms the WFST decoder when applied to the Transformer on both benchmarks ($p < 0.05$; paired t-test across $n = 10$ seeds). Furthermore, RTF values are lower for both decoders with the Transformer (Table~\ref{tab:transformer}).

\begin{table}[b!]
  \caption{Ablation study on B2T '24 and B2T '25 validation sets with baseline GRU. Values are Mean WER (\%) $\pm$ SEM ($n=10$ seeds). $\dagger$ denotes significant degradation relative to Full model.}
  \label{tab:ablation_flat}
  \centering
  \begin{tabular}{ l r@{}l r@{}l }
    \toprule
    \textbf{Configuration} & \multicolumn{2}{c}{\textbf{B2T '24 WER}} & \multicolumn{2}{c}{\textbf{B2T '25 WER}} \\
    \midrule
    \textbf{LightBeam (Full)}       & 14&.17 \scriptsize{$\pm$ 0.08}            & 6&.36 \scriptsize{$\pm$ 0.05} \\
     No LLM first-pass rescore& 17&.17$^\dagger$ \scriptsize{$\pm$ 0.05}    & 8&.31$^\dagger$ \scriptsize{$\pm$ 0.07} \\
     No NWP Fine-tuning    & 15&.89$^\dagger$ \scriptsize{$\pm$ 0.09}    & 7&.26$^\dagger$ \scriptsize{$\pm$ 0.04} \\
    No Phoneme Variants   & 14&.05 \scriptsize{$\pm$ 0.07}            & 6&.36  \scriptsize{$\pm$ 0.05} \\
    \bottomrule
  \end{tabular}
\end{table}

We next compared the performance of applying generative error (GEC) correction with the time-masked Transformer when using either the WFST decoder or \textit{LightBeam} (Section 
\ref{sec:generative}). We used Llama 3.1 8B for GEC. We include GEC results with both a causal and non-causal time-masked Transformer, because Zhang et al. \cite{zhang2026decoding} achieved SOTA results ($5.10\%$ on B2T '24, $2.21\%$ on B2T '25 public) by pretraining a noncausal Transformer on 367 hours of intracranial data and using GPT 3.5 or GPT 4 for GEC. For both versions of the Transformer,  GEC with \textit{LightBeam} performed significantly better than with the WFST decoder on B2T '24 and B2T '25 public ($p < 0.05$; paired t-test across $n=10$ seeds) and there was no significant difference on B2T '25 private (Figure \ref{fig:generative_error_corection}).  The mean WER across seeds with the noncausal Transformer was $5.01\%$, $2.36\%$, and $2.08\%$ on B2T '24 and B2T '25 private and public, respectively  (Figure \ref{fig:generative_error_corection}). These results are significantly better than those of Zhang et al. ($p < 0.05$; one-sample t-test), establishing a new SOTA relative to published results. Notably, \textit{LightBeam} does not rely on extensive pretraining or large close-sourced LLMs. Importantly, \textit{LightBeam} innovates on the decoder while the performance gains from Zhang et al. stem from changes to the encoder, and so these research directions are orthogonal and complementary. 

\begin{figure}[t!]
  \centering
  \includegraphics[width=\linewidth, height=4cm, keepaspectratio]{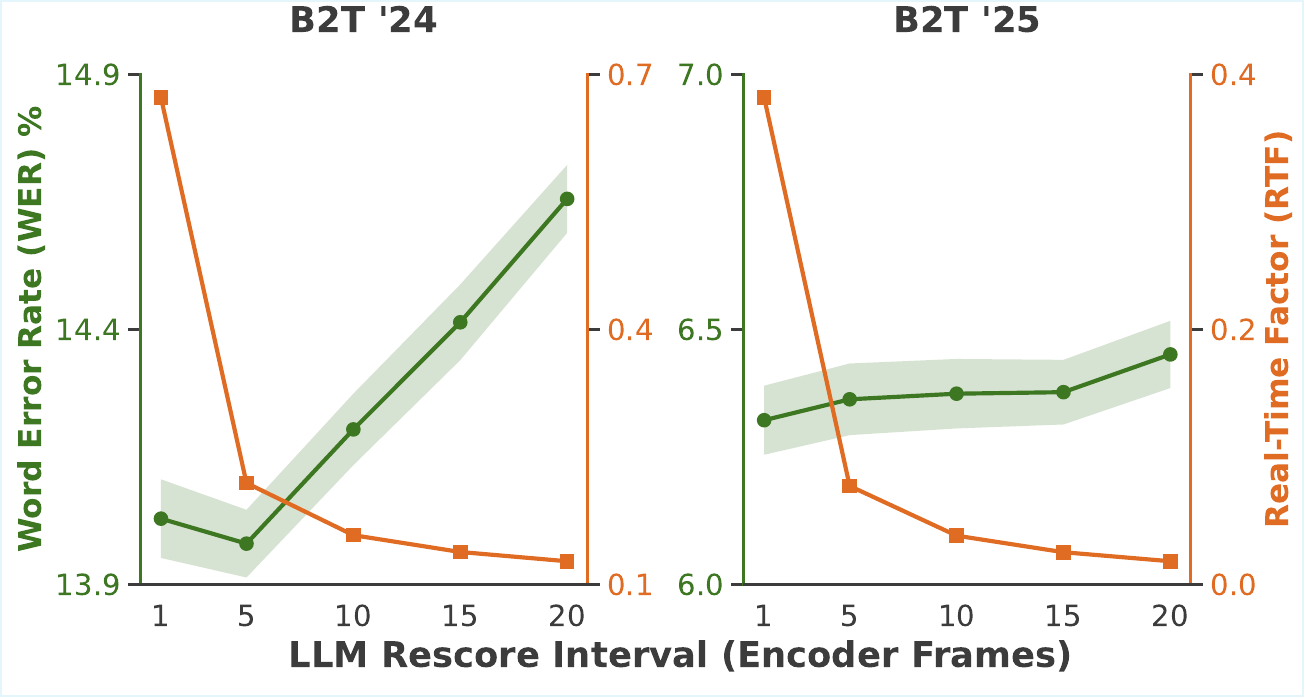}
  \caption{Impact of LLM rescore interval on validation WER and RTF across both datasets. An encoder frame is output every 100 ms in B2T '24, and every 80 ms in B2T '25. Shaded regions indicate SEM across $n=10$ seeds. LLM rescore interval was set to 10 and 15 throughout the study for B2T '24 and '25, respectively.}
  \label{fig:llm_rescore_interval}
\end{figure}

\subsection{Ablations and hyperparameter sweeps}
To isolate the impact of \textit{LightBeam}'s core components, we conducted an ablation study on the B2T '24 and '25 validation sets using the baseline GRU (Table~\ref{tab:ablation_flat}). First, we evaluated the necessity of delayed fusion by reverting to a traditional pipeline where the LLM is applied only once at the end of the trial. Removing the continuous LLM rescoring—which occurs at 1~s and 1.2~s intervals for B2T '24 and '25—resulted in significant performance degradation across both datasets (Table \ref{tab:ablation_flat}), underscoring the advantage of integrating the LLM directly into the first-pass beam search. Next, we ablated the LLM next-word prediction (NWP) fine-tuning; removing this adaptation also caused significant degradation, highlighting the importance of tailoring the LLM to the participant's specific text distribution. Finally, we assessed whether permitting multiple phonetic pronunciations per word (Supplementary Section I) improved performance; we found that using only default pronunciations did not result in significant degradation. We further analyzed the impact of the LLM rescore interval (Figure \ref{fig:llm_rescore_interval}); increasing the interval generally increased WER while decreasing RTF, demonstrating a tunable tradeoff between accuracy and latency.

\section{Discussion}
In this study, we present \textit{LightBeam}, a memory-efficient CTC decoder for speech neuroprostheses that establishes a new SOTA on the B2T '24 and '25 benchmarks with real-time decoding speeds. \textit{LightBeam} achieves these holistic gains  by integrating an LLM into first-pass decoding, reducing reliance on a large N-gram LM. We identify three primary limitations. First, while \textit{LightBeam} exhibits higher latency than the WFST baseline, its processing speed on remains within the constraints required for real-time decoding (RTF $<$ 1). Second, because LLM scores are integrated at fixed intervals, significant divergence between the LLM and N-gram LM scores may cause a degree of  ``jitter'' in the decoded output. Third, the reliance on next-word prediction fine-tuning implies that \textit{LightBeam} may require an initial calibration phase to adapt to the idiosyncratic text distributions of new participants. 

Future research could explore several promising directions. First, it would be valuable to observe if continuous next-word prediction fine-tuning allows \textit{LightBeam} to adapt to a patient’s evolving linguistic nuances over long-term use. Second, given that LLMs excel at processing long-range dependencies, \textit{LightBeam} may provide even greater performance gains in long-context conversational trials. Finally, the modular nature of delayed fusion allows for the integration of other LLMs, potentially improving accuracy further as open-source models improve (we report limited results with Llama 3.2 3B Base in Supplementary Section J). By providing an open-source Python implementation, we hope to enable the research community to build upon \textit{LightBeam} to make speech neuroprostheses more accessible and accurate for patients.

\clearpage

\section{Acknowledgements}
This work was supported by the following awards. To : NSF CAREER 1943467, NIH DP2NS122037, NIH R01NS121097. To EF: NIH T32NS115753.

\section{Generative AI Use Disclosure}
We used Claude to assist with the implementation of \textit{LightBeam} and all AI-assisted edits were verified by the researchers. We used Gemini to assist with writing this manuscript to improve grammar and diction. 

\section{Author Contributions}
\textbf{E.F.}: Conceptualized the \textit{Lightbeam} decoder framework, developed the core software implementation, and authored the initial manuscript draft. 
\textbf{J.H.}: Implemented the Transformer training pipeline with limited left-context constraints, helped author supplementary materials, generated the Figures, and provided revisions to the manuscript. 
\textbf{N.H.}: Contributed to the conceptualization of the Lightbeam architecture and provided revisions to the manuscript. 
\textbf{J.C.K.}: Supervised the research, acquired the necessary funding, and provided revisions to the manuscript.

\bibliographystyle{IEEEtran}
\bibliography{mybib}

\end{document}


\appendix

\section{Additional benchmark details}
\subsection{Brain-to-Text 2024}
B2T '24 is partitioned into a training set of 8,800 trials, a validation set of 880 trials, and a test set of 1,200 trials (for which ground-truth sentence labels are withheld). Trials are organized into blocks, with each block containing 20--50 sentences. Specifically, the validation split consists of the final block from 22 out of the 24 total recording days. The test split comprises the first two blocks from 15 of the 24 recording days. The training split contains all remaining blocks across the full 24-day recording period. All sentences in this benchmark are sourced exclusively from the Switchboard corpus. All data is publicly available on dryad, with preprocessing details provided in Willett et al. \cite{Willett2023-an}.

\subsection{Brain-to-Text 2025}
B2T '25 is partitioned into a training set of 8,072 trials, a validation set of 1,426 trials, and a test set of 1,450 trials (for which ground-truth sentence labels are withheld). Trials are grouped into blocks of 5--50 sentences. The validation and test sets combined comprise 67 blocks collected across 41 of the 45 recording days, while the training set comprises 198 blocks spanning all 45 days. The distribution of corpus sources across these splits is detailed in Table \ref{tab:data_distribution}. The linguistic material is sourced from a diverse set of corpora, including Switchboard, OpenWebText2, Harvard Sentences,  a custom high-frequency word list, and random word sequences. All data is publicly available on dryad, with preprocessing details provided in Card et al. \cite{Card2024-an}
\begin{table}[ht]
    \centering
    \caption{Distribution of data sources across training and evaluation splits.}
    \label{tab:data_distribution}
    \small
    \setlength{\tabcolsep}{8pt}
    \begin{tabular}{l ccc c}
        \toprule
        \textbf{Source} & \textbf{Blocks} & \textbf{Sentences} &  \textbf{Split} \\
        \midrule
        Switchboard & 182 & 7,672  & Train, Val/Test \\
        Freq. Words & 48  & 2,050  & Train, Val/Test \\
        50-word Vocab & 25  & 1,091  & Train Only \\
        Openwebtext	& 5	& 172 & Val/Test only \\
        Harvard	& 3	& 98 &	Val/Test only \\
        Random	& 2	& 79 &	Val/Test only \\
        \bottomrule
    \end{tabular}
\end{table}

\section{CTC encoder hyperparameters}

For the GRU encoders, we follow the exact same implementation of Willett et al. \cite{Willett2023-an} for B2T '24 and Card et al.\ \cite{Card2024-an} for B2T '25. For the Transformer encoders, we adopted the parameters from Feghhi et al.\ \cite{feghhi2025time} for B2T '24. The hyperparameters for the B2T '25 time-masked Transformer are listed in Table \ref{tab:transformer_params}.

\begin{table}[h]
  \caption{Transformer architecture and training hyperparameters for B2T '25. Feed-forward network (FFN) multiplier indicates the increase in dimensionality in the FFN module.}
  
  \label{tab:transformer_params}
  \centering
  \begin{tabular}{l c} 
    \toprule
    \textbf{Hyperparameter} & \textbf{Value} \\ 
    \midrule
     Patch Length & 4 (0.08 ms)  \\
     Patch Width & 512 \\
     Transformer Dimension & 384\\
     Layers & 5 \\
     Heads & 6\\
     Head Dimension & 64\\
     FFN Multiplier & 4 \\
     Learning Rate & 0.002 \\
     White Noise & 0.2\\
     Baseline Shift & 0.05\\
     Number of Time-Masks & 20\\
     Max Mask Length & 5\\
     Epochs & 500 \\
     Dropout & 0.35\\
     Input Dropout & 0.2\\
     L2 Decay & $1 \times 10^{-5}$\\
     Gaussian Smooth Kernel Size & 20 \\
     Gaussian Smooth Width & 2.0 \\
     Gradient Norm Clipping & 100\\
    \bottomrule
  \end{tabular}
\end{table}

\section{Bounded KV Cache}
While GRU architectures compress the input into a fixed hidden state, Transformers require access to all previous input, resulting in an unbounded Key-Value (KV) cache. This is problematic when deploying the speech neuroprostheses in an unstructured, conversational setting where the context is not reset after each sentence. Inspired by Zheng et al.\cite{Zheng2023-an}, we limit the left context of the causal Time-Masked Transformer to guarantee a constant Key-Value (KV) cache, making the Transformer suitable for real-world, clinical applications.  

\subsection{Dynamic left temporal context constraint}
We define the input sequence as $X = \{x_1, x_2, \dots, x_T\}$, where $T$ is the total sequence length. In a standard causal Transformer, the attention at time step $t$ is computed over all $x_i$ where $i \leq t$. Our model modifies the attention mask $M \in \{0, 1\}^{T \times T}$ such that:$$M_{t,i} = \begin{cases}1, & \text{if } \max(0, t - \tau) \leq i \leq t \\ 0, & \text{otherwise}\end{cases}$$where $\tau$ represents the temporal lookback window. During training, $\tau$ is sampled stochastically from a uniform distribution $\tau \sim \mathcal{U}(\tau_{min}, \tau_{max})$, where $\tau_{min} = 1\text{s}$ and $\tau_{max} = 20\text{s}$. By dynamically sampling $\tau$ during training, the Transformer can operate with a variety of left context sizes during inference.

\subsection{KV cache space complexity}
Let $d$ be the model dimension and $L$ the number of layers. In a standard causal Transformer, the memory requirement $S$ at time $t$ grows linearly with respect to time for self-attention layers:$$S_{standard}(t) = t \cdot d \cdot L \cdot 2$$
When restricting left context, the memory requirement is bounded by the maximum context window $\tau_{max}$. Let $f_s$ be the sampling frequency of  the encoder (which is the inverse of the encoder frame length). The maximum number of frames stored, $N_{max}$, is defined as $N_{max} = \tau_{max} \cdot f_s$. The space complexity per inference step is:$$S_{constrained}(t) = \min(t, N_{max}) \cdot d \cdot L \cdot 2$$As $t \to \infty$ (streaming scenarios), the memory footprint remains $O(1)$ relative to the total sequence length $T$, whereas the standard self-attention computation remains $O(T)$.

\subsection{Computational efficiency and stability}
Beyond memory, this constraint also ensures constant-time inference per token. In global attention, the $t$-th token requires $O(t)$ FLOPs for the attention dot-product, leading to a total quadratic complexity of $O(T^2)$. By enforcing the left-context limit, the per-step complexity is capped:$$\text{Attention FLOPs per step} \approx O(N_{max} \cdot d)$$This results in a total complexity of $O(T \cdot N_{max})$, which is strictly linear with respect to the input duration. 

\subsection{Performance}
In our experiment, we examined choices of different left context size and their impact on encoder performance measured in Phoneme Error Rate (PER).
\begin{figure}[h!]
    \centering
    \includegraphics[width=0.7\linewidth]{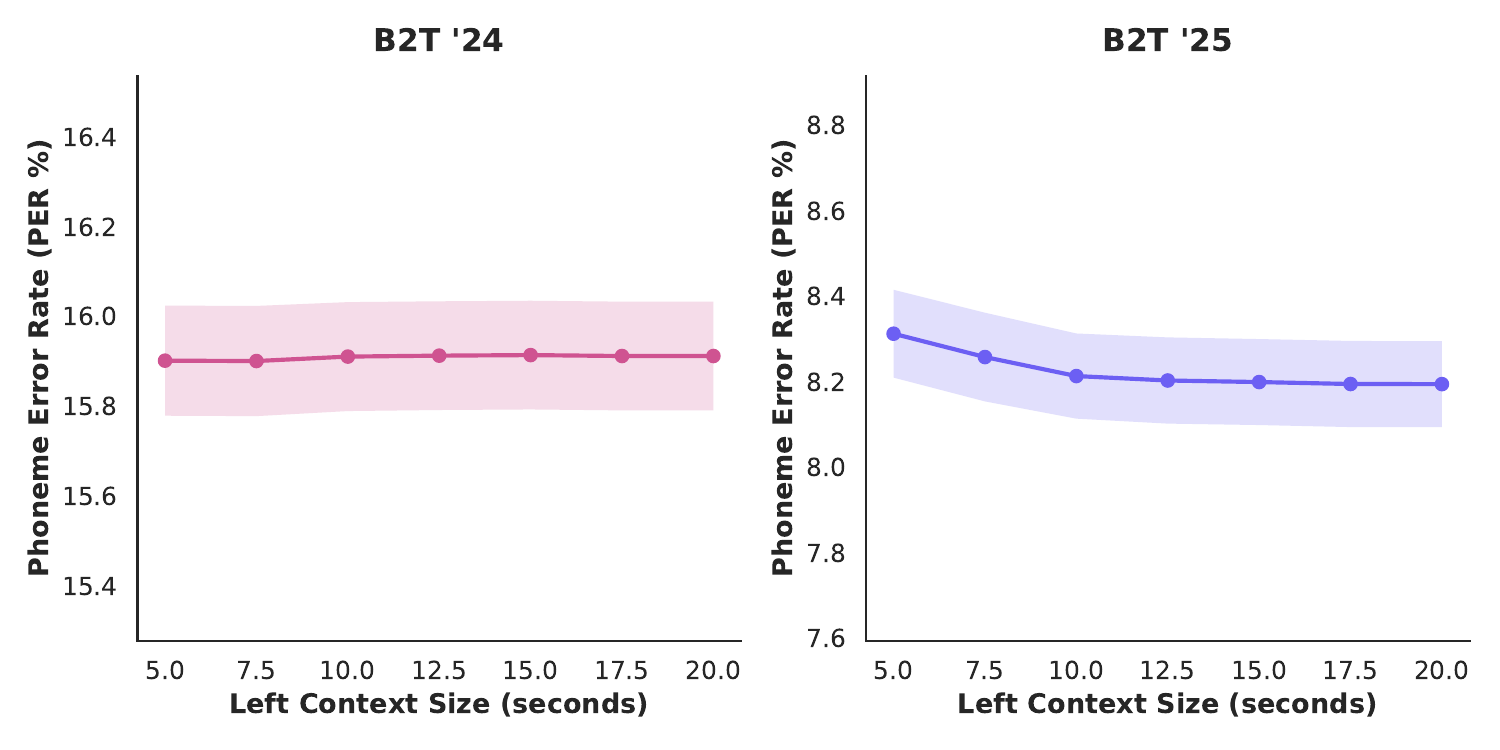}
    \caption{Phoneme Error Rate (PER) as a function of left-context window size for B2T'24 and B2T '25 on validation set.}
    \label{fig:left_context_per}
\end{figure}
We show in Figure \ref{fig:left_context_per} that encoder performance across left-context window sizes remains stable. We used a left context size of 7.5 seconds for B2T '24  and 20 seconds for B2T '25 throughout the paper.

\section{WFST decoder hyperparameters and additional details}

Hyperparameters for the WFST decoder are listed in Table \ref{tab:WFST_first_pass} and \ref{tab:wfst_2nd_pass}. For additional details on these hyperparameters and additional implementational detail refer to Willett et al. \cite{Willett2023-an}.

\begin{table}[h!]
  \caption{WFST decoding hyperparameters for B2T '24 and '25 benchmarks during first-pass rescoring. These hyperparameters were tuned by the original authors using the baseline GRU encoder, and the same set of hyperparmeters were used for analyses with the time-masked Transformer. Variable notation is copied from original authors (\cite{Willett2023-an, Card2024-an}).}
  \vspace{5pt}
  \label{tab:WFST_first_pass}
  \centering
  \small
  \begin{tabular}{l cc} 
    \toprule
    \textbf{Hyperparameter} & \textbf{B2T' 24} & \textbf{B2T' 25} \\ 
    \midrule
    Blank Penalty& log(7)& log(90)\\    
    Acoustic Scale ($\alpha$)& 0.5& 0.325\\
    Beam& 18& 17\\
    Max Active States& 7000& 7000\\
    Min Active States& 200& 200\\
    \bottomrule
  \end{tabular}
\end{table}

\begin{table}[h!]
  \caption{WFST decoding hyperparameters for B2T'24 and '25 benchmarks during second-pass rescoring. These hyperparameters were tuned by the original authors using the baseline GRU encoder, and the same set of hyperparmeters were used for analyses with the time-masked Transformer.Variable notation is copied from original authors (\cite{Willett2023-an, Card2024-an}).}
  \vspace{5pt}
  \label{tab:wfst_2nd_pass}
  \centering
  \small
  \begin{tabular}{l cc} 
    \toprule
    \textbf{Hyperparameter} & \textbf{B2T' 24} & \textbf{B2T' 25} \\ 
    \midrule
    Acoustic Scale ($\alpha$)& 0.5& 0.325\\  
    N-gram LM weight ($\beta$)& 0.5& 0.45\\
    LLM weight& 0.5& 0.55\\
    N-best beam size& 100& 100\\
    Used quantized LLM& Yes& No\\
    \bottomrule
  \end{tabular}
\end{table}

\section{LightBeam hyperparameters and additional details}

The full list of hyperparameters for \textit{LightBeam} are listed in Table \ref{tab:hparams_lightbeam}. Additional technical details are provided below.

\label{appendix:decoder_hyperparams}
\begin{table}[h!]
  \caption{\textit{LightBeam} decoding hyperparameters for B2T '24 and B2T '25. These hyperparameters were tuned using the GRU encoder, and the same set of hyperparmeters were used for analyses with the time-masked Transformer.}
  \vspace{5pt}
  \label{tab:hparams_lightbeam}
  \centering
  \small
  \begin{tabular}{l cc} 
    \toprule
    \textbf{Hyperparameter} & \textbf{B2T' 24} & \textbf{B2T' 25} \\ 
    \midrule
    LLM rescore interval (Encoder Frames)       & 10    & 15    \\
    LLM Weight ($\phi$) & 1.2   & 1.2   \\    
    N-gram LM Weight ($\omega$)& 0.8   & 1.0   \\
    Acoustic Scale ($\alpha$)            & 0.6   & 0.4   \\
    Beam Size (\textit{k})& 1000  & 900   \\
    Beam Prune Threshold ($\theta$)& 22.0  & 18.0  \\
    Homophone Beam Number (\textit{h})& 3     & 3     \\
    Homophone Prune Threshold ($\lambda$) & 4.0   & 4.0   \\
    Token Insertion Bonus ($\beta$)& 1.5   & 1.5   \\
    Word Boundary Bonus ($\gamma$)& 1.0   & 1.0   \\
    LLM Scoring Chunk Size        & 256   & 256   \\
    \bottomrule
  \end{tabular}
\end{table}

\subsection{Beams object}

The beam search state is encapsulated in a \texttt{Beams} object whose tensors are pre-allocated on the GPU before decoding begins. The object maintains, for each of the $K$ active 
hypotheses, a cumulative log-probability score $\mathbf{s} \in \mathbb{R}^{K}$, the most recently emitted non-blank token $\boldsymbol{\ell}_{\text{last}} \in \mathbb{Z}^{K}$, and a rolling
hash $\mathbf{h} \in \mathbb{Z}^{K}$ of the decoded token sequence. Rather than storing $K$ independent token sequences of length up to $T$, hypotheses are stored as an implicit tree: a label
buffer $\mathbf{W} \in \mathbb{Z}^{K \times T}$ records the token emitted at each step, and a companion pointer buffer $\mathbf{P} \in \mathbb{Z}^{K \times T}$ records which parent beam each
hypothesis branched from, allowing the full sequence of any hypothesis to be reconstructed by following pointers in reverse. After each decoding step, hypothesis recombination merges beams
that have converged to the same decoded text — detected in $\mathcal{O}(K)$ time via hash comparison rather than $\mathcal{O}(K^2)$ pairwise string equality — with duplicate scores
consolidated via a \texttt{max} reduction. Because all buffers are pre-allocated and reused across frames, no dynamic GPU memory allocation occurs during the decoding loop. While the current work decodes single utterances ($K$ beams), the data structure natively supports larger batch sizes by extending all tensors along a batch dimension, enabling
straightforward scaling to higher-throughput settings.

\subsection{Vectorized Lexicon}
Enforcing a lexicon constraint during CTC beam search requires, at each decoding step, determining the set of valid next tokens for every active beam hypothesis. A natural implementation
represents the lexicon as a prefix tree (trie), where each node encodes the valid continuations of a phoneme prefix. Given a beam of $k$ hypotheses, the trie-based approach must traverse each
hypothesis path independently on the CPU, performing $\mathcal{O}(l)$ pointer-chasing dictionary lookups per hypothesis for a prefix of length $l$, yielding a total cost of $\mathcal{O}(kl)$
sequential operations per decoding step. Beyond raw complexity, this approach introduces a critical synchronization bottleneck: the constraint mask must be assembled on the CPU and
transferred to the GPU before each log-probability masking step, stalling the GPU pipeline.

To eliminate this bottleneck, we flatten the trie into a dense \emph{transition table} $\mathbf{T} \in \mathbb{Z}^{S \times V}$, where $S$ is the number of trie nodes and $V$ is the
vocabulary size. The entry $\mathbf{T}[s, v]$ stores the successor state reached from prefix state $s$ upon observing token $v$, or a designated sink state $s_\varnothing$ for invalid
transitions. The prefix state is represented as a single integer tensor $\mathbf{s} \in \mathbb{Z}^{K}$ residing on the GPU. At each step, computing the constraint mask reduces to a single
gather operation, $\mathbf{M} = (\mathbf{T}[\mathbf{s}] \neq s_\varnothing) \in {0,1}^{K \times V}$, which executes across all $K$ hypotheses simultaneously. Two special cases follow directly
from CTC semantics: the blank token is always unmasked, and the most recently emitted non-blank token is always permitted as a repeat. In both cases the prefix state $\mathbf{s}$ is left
unchanged, since neither blanks nor repeated tokens advance the decoded prefix. Advancing the prefix state is therefore guarded by $\boldsymbol{\ell} \neq \text{blank}$ and $\boldsymbol{\ell}
\neq \boldsymbol{\ell}_{\text{prev}}$, after which the update reduces to a single vectorized lookup $\mathbf{s}' = \mathbf{T}[\mathbf{s},, \boldsymbol{\ell}]$. By keeping all constraint
state and computation resident on the GPU, the vectorized lexicon constraint removes the CPU--GPU synchronization overhead entirely and scales gracefully with larger beam widths.

\subsection{Homophone tracking and shallow fusion with N-gram LM}
Because phoneme-based CTC decoding is agnostic to orthography, a single acoustic beam hypothesis may correspond to multiple valid word spellings (homophones, e.g., \textit{aunt} /            
\textit{ant}). To handle this, each of the $K$ acoustic beams maintains $o$ orthographic sub-beams — candidate word-level text hypotheses that share the same underlying phoneme path. Each    
orthographic sub-beam tracks a cumulative N-gram LM score, an integer state ID into the N-Gram LM state registry, and a backpointer into a shared \texttt{WordHistory} trie that reconstructs full word sequences without redundant string allocation. Word-level scoring is triggered precisely when a beam emits the word boundary token: the vectorized lexicon state is     
queried to retrieve all lexicon words that complete at the current prefix, and the N-gram LM is evaluated for each (orthographic context, candidate word) pair. Transition scores are cached by
(state ID, word) key so that each unique context--word pair calls the N-gram LM at most once across the entire beam and across utterances. The weighted LM score $\omega \cdot \log P_{\mathcal{G}}(w
\mid \text{context})$ is accumulated into each orthographic sub-beam's running total, and the acoustic beam score is updated by the delta between the new best orthographic score and the
previous best.

After scoring, orthographic sub-beams are ranked and the top $o$ candidates are retained. Any candidate whose cumulative LM score falls more than $\lambda$ log-probability points below the
best surviving candidate is discarded, preventing low-probability homophones from inflating memory and compute. The acoustic beam score therefore always reflects the best available word-level
interpretation, while the full ranked list of orthographic sub-beams is preserved for subsequent neural LM rescoring. The N-gram LM is loaded and queried via Ken-LM \cite{heafield2011kenlm}.

\subsection{Delayed fusion with LLM}
While the N-gram LM provides efficient word-boundary scoring during the decoding loop, an LLM is additionally used to rescore the $o$ orthographic sub-beams of each acoustic hypothesis at
regular intervals during decoding and once more at end-of-sentence. The two models play complementary roles: the N-gram LM is well-suited to the word-boundary expansion step, where the number
of orthographic candidates is variable and potentially large due to homophones — its $\mathcal{O}(1)$ cached state transitions make it cheap to score every candidate word in every context.
The LLM, by contrast, requires a full forward pass per hypothesis and is therefore best applied when the number of orthographic sub-beams is fixed at $o$, amortizing its cost over a known,
bounded set of candidates. At each rescoring event, the full word sequences of all orthographic sub-beams across all $k$ acoustic beams are collected and deduplicated, so that identical
hypotheses appearing in multiple beams are scored only once. Each unique text is scored by computing $\sum_t \log p_{\text{LLM}}(w_t \mid w_{<t})$, and the resulting score replaces the N-gram
score for each orthographic sub-beam. The beam score is then updated by the delta between the new best and old best orthographic sub-beam score.

At the end of decoding, an end-of-sentence scoring pass appends each of the candidate punctuation tokens ${\texttt{.},\ \texttt{?},\ \texttt{!}}$ to every orthographic sub-beam, scores each
variant through the LLM, and selects the punctuation that maximizes $\log p_{\text{LLM}}$. This allows the system to jointly infer sentence-final punctuation and refine hypothesis ranking
based on full-sequence LLM probability, rather than committing to punctuation as a post-processing step. The result is broadcast back to all requesting sub-beams, ensuring the number of LLM
forward passes scales with the number of \emph{unique} hypotheses rather than $k \times o$.

\section{Next-word prediction finetuning}
Next-word prediction finetuning is performed using the training sentences across both datasets. Each sentence is converted to sentence-case because the N-gram LM is not cased. LoRA adapters are applied to all attention projection layers ($W_q, W_k, W_v, W_o$) and MLP layers with rank $r=16$ and $\alpha=32$ \cite{Hu2022-ph}. The model is trained for 1 epoch using the AdamW optimizer with learning rate $2\times10^{-4}$ and weight decay $10^{-3}$,  with a batch size of 16 in BFloat16 precision. Hyperparameters were tuned using the validation sentences from both datasets, and are listed in Table \ref{tab:params_delayed_fusion}.

\begin{table}[ht]
    \centering
    \caption{Fine-tuning hyperparameters for the next-word prediction task}
    \vspace{5pt}
    \label{tab:params_delayed_fusion}
    \small
    \begin{tabular}{l c}
        \toprule
        \textbf{Hyperparameter} & \textbf{Value} \\
        \midrule
        Base Precision & BFloat16 \\
        Optimizer & AdamW \\
        Learning Rate & $2 \times 10^{-4}$ \\
        Weight Decay & $10^{-3}$ \\
        Batch Size & 16 \\
        Epochs & 3 \\
        Eval. Frequency & 0.25 Epoch \\
        \midrule
        \textit{LoRA Configuration} & \\
        Rank ($r$) & 16 \\
        Alpha ($\alpha$) & 32 \\
        Target Modules & $W_q, W_k, W_v, W_o$ \\
        \bottomrule
    \end{tabular}
\end{table}

\section{Generative error correction}
We adapt the procedure from Feghhi et al. \cite{feghhi2025time} for generative error correction. To perform LLM fine-tuning, we first generated decoded transcripts using 10 seeds of the time-masked Transformer using the WFST decoder and \textit{Lightbeam}
separately. Decoded transcripts were generated by taking the top-scoring beam from each seed across the training, validation, and testing sets. We then fine-tuned Llama 3.1 8B to predict the ground-truth transcript given the 10 decoded sentences for each trial. Fine-tuning was performed by ``masking'' the ground-truth transcript and training the LLM to predict the masked tokens with cross-entropy loss. The LLM was fine-tuned with QLoRA \cite{Dettmers2023-hl} via the Unsloth package \cite{unsloth}. We used the default hyperparameters provided by Unsloth for Llama 3.1 8B, which are listed in Table \ref{tab:params_corrector}. A prompt, also provided below, was used to guide the LLM for each trial. As no hyperparameter tuning was performed, our procedure consisted of fine-tuning the LLM first on the validation set, then on the training set, and finally once more on the validation set. The validation and training sets were combined across B2T '24 and '25. We report results across 10 seeds of fine-tuning.

\begin{table}[ht]
    \centering
    \caption{Fine-tuning hyperparameters for the generative error correction}
    \vspace{5pt}
    \label{tab:params_corrector}
    \small
    \begin{tabular}{l c}
        \toprule
        \textbf{Hyperparameter} & \textbf{Value} \\
        \midrule
        Quantization & 4-bit NF4 (bitsandbytes) \\
        Optimizer & AdamW-8bit \\
        Learning Rate & $2 \times 10^{-4}$ \\
        Batch Size & 16 \\
        Epochs & 1 \\
        Data Scheduling & \texttt{[val, train, val]} \\
        Acceleration Framework & Unsloth \\
        \midrule
        \textit{LoRA Configuration} & \\
        Rank ($r$) & 16 \\
        Alpha ($\alpha$) & 16 \\
        Target Modules & $W_q, W_k, W_v, W_o$, MLP \\
        \bottomrule
    \end{tabular}
\end{table}

\begin{tcolorbox}[
    enhanced,
    colback=gray!5,
    colframe=gray!50,
    width=\linewidth,      
    sharp corners,         
    boxrule=0.5pt,
    left=10pt, right=10pt, 
    top=8pt, bottom=8pt,
    fontupper=\small\ttfamily,
    before upper=\justifying\let\\\newline 
]
Your task is to perform automatic speech recognition error correction. Below are multiple candidate transcriptions of the same utterance. These candidates were decoded from neural activity and may contain errors. Based on the candidates, produce the single most accurate, coherent, and grammatical transcription of the original utterance. Focus on key differences between candidates that change meaning or correctness, and avoid repetitive or nonsensical phrases. Respond with only the final corrected transcription--no explanations or extra text.
\end{tcolorbox}

\section{Real-Time Factor (RTF)}
\begin{figure}[h!]
    \centering
    \includegraphics[width=0.8\linewidth]{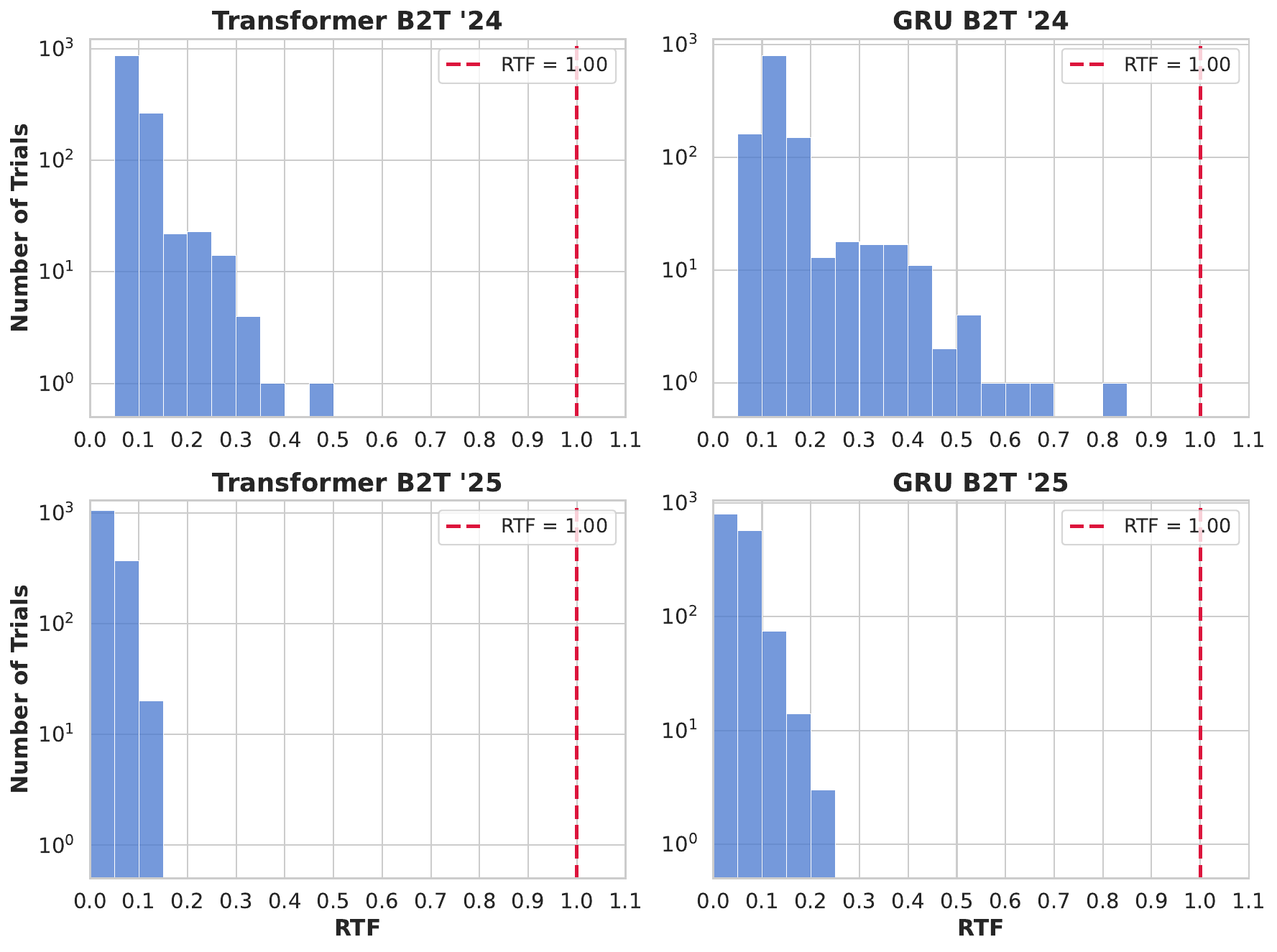}
    \caption{RTF Distributions of \textit{LightBeam} across all trials on the test set of B2T '24 and B2T '25. The scale of the y-axis is logarithmic.}
    \label{fig:rtf}
\end{figure}

Real-Time Factor (RTF) is a common metric for measuring the latency of sequence processing systems. Given a system which takes time $f(l)$ to process an input sequence of duration $l$, the RTF is defined as: $$\mathrm{RTF} = f(l)/l$$
Decoding with \textit{LightBeam} achieves RTF $<1.00$ regardless of encoder model choice on the test set, which makes the decoding pipeline ideal for clinical application (Figure \ref{fig:rtf}). 

\section{Multiple pronunciation variants}
We constructed a pronunciation lexicon from the 100k-word vocabulary of the 4-gram language model provided by ImagineVille. Each word was paired with its default CMU pronouncing dictionary phoneme sequence, and words with multiple CMU pronunciations received additional entries in the lexicon  (denoted word(2), word(3), etc.) for each distinct variant. Trailing numbers were stripped from words before applying the N-gram LM for scoring. We found that allowing for alternative pronunciations of the same word did not significantly improve performance over just using the default pronunciation for each word. 

\section{Llama 3.2 3B Base for delayed fusion}
We experimented using a finetuned Llama-3.2-3B in place of the finetuned Llama-3.2-1B for delayed fusion on the validation set for the baseline GRU. We observed statistically significant improvement in performance on B2T '24, but no significant improvement for B2T '25, at the cost of higher RTF and VRAM usage (Table \ref{tab:transformer}).

\begin{table}[h!]
  \caption{Word Error Rate (\%) comparison on the B2T '24 and B2T '25 Validation set when using GRU encoder and different LLMs for Delayed Fusion. Mean $\pm$ SEM over $n=10$ seeds. ($^\dagger$ sig. better than 1B Model).}
  \vspace{5pt}
  \label{tab:transformer}
  \centering
  \footnotesize
  \setlength{\tabcolsep}{4.5pt} 
  \begin{tabular}{l | ccc | ccc}
    \toprule
    \multirow{2}{*}{\textbf{LLM}} & \multicolumn{3}{c|}{\textbf{B2T '24}} & \multicolumn{3}{c}{\textbf{B2T '25}} \\
    \cmidrule(lr){2-4} \cmidrule(lr){5-7}
    & \textbf{WER} & \textbf{RTF} & \textbf{VRAM} & \textbf{WER} & \textbf{RTF} & \textbf{VRAM} \\
    \midrule
    llama-3.2-1B & 14.17{\scriptsize$\pm$.22} & 0.18 & 6.14 & 6.36{\scriptsize$\pm$.14}   & 0.06 & 5.60\\
    llama-3.2-3B & \textbf{13.82$^\dagger${\scriptsize$\pm$.13}} & 0.26 & 9.82  &  6.24{\scriptsize$\pm$.16} & 0.08 & 9.64\\
    \bottomrule
  \end{tabular}
\end{table}

\section{Hyperparameter tuning}
All hyperparameter tuning was done using the validation sets. For the time-masked Transformer Transformer on B2T '25, we performed tuning by starting with the hyperparameters of the B2T '24 model and modifying them manually based on validation phoneme error rate. For \textit{LightBeam}, we performed all hyperparameter tuning manually using the baseline GRU. 

\clearpage

\bibliographystyle{plain}
\bibliography{mybib}